\documentclass[prd,onecolumn,preprintnumbers,nofootinbib]{revtex4}
\usepackage{graphicx}

\usepackage{color}
\usepackage{float}
\newcommand{\rthis}[1]{\textcolor{black}{#1}}
\usepackage{amsfonts,amsmath,amssymb}
\usepackage[plainpages=false, colorlinks=true, anchorcolor=blue, linkcolor=blue, citecolor=blue, bookmarks=false]{hyperref}
\usepackage{natbib}
\usepackage{enumitem}
\usepackage{lipsum}
\usepackage{multirow}
\usepackage{graphicx}
\usepackage{subfigure}
\usepackage{array}
\pdfoutput=1
\begin{document}
%\markboth{Authors' Names}
%\newcommand{\mthis}[1]{\textcolor{red}{#1}}
\makeatletter
\newcommand{\thickhline}{%
    \noalign {\ifnum 0=`}\fi \hrule height 1pt
    \futurelet \reserved@a \@xhline
}
\newcolumntype{"}{@{\hskip\tabcolsep\vrule width 1pt\hskip\tabcolsep}}
\makeatother
\providecommand{\brak}[1]{\ensuremath{\left(#1\right)}} %for longer brackets
\newcommand{\bthis}[1]{\textcolor{black}{#1}}
\newcommand{\apjl}{Astrophys. J. Lett.}
\newcommand{\apjs}{Astrophys. J. Suppl. Ser.}
\newcommand{\aap}{Astron. \& Astrophys.}
\newcommand{\nar}{New  Astronomy Reviews}
\newcommand{\aapr}{Astron and Astrophys Reviews}
\newcommand{\aj}{Astron. J.}
\newcommand{\araa}{Ann. Rev. Astron. Astrophys. } %ARA$\&$A}
\newcommand{\mnras}{Mon. Not. R. Astron. Soc.}
\newcommand{\ssr}{Space Science Revs.}
\newcommand{\apss}{Astrophysics \& Space Sciences}
\newcommand{\jcap}{JCAP}
\newcommand{\pasj}{PASJ}
\newcommand{\pasp}{PASP}
\newcommand{\pasa}{Pub. Astro. Soc. Aust.}
\newcommand{\physrep}{Phys. Rep.}
\renewcommand{\arraystretch}{2.5}
\setlength{\tabcolsep}{12pt}

\title{ A test of linearity of the  ratio of  dark matter to baryonic matter  in galaxy clusters}
\author{Varenya \surname{Upadhyaya}}\altaffiliation{E-mail:ep20btech11026@iith.ac.in}

\author{Shantanu  \surname{Desai}}  
\altaffiliation{E-mail: shntn05@gmail.com}

\begin{abstract}
We search for a linearity in the ratio of  dark matter to baryonic matter as a function of radius for galaxy clusters, motivated by a recent result by Lovas (2022), who has discovered such a linearity for a diverse suite of galaxies in the SPARC sample. For our analysis, we used a sample of 54 non-cool core clusters from the HIFLUGCS sample.   We do not find any evidence for a linear trend in the aforementioned ratio as a function of radius for individual clusters. We then repeat this analysis for the stacked sample, which also does not show this linearity. Therefore, the linear scaling found by Lovas is not a universal property of dark matter haloes at all scales.
\end{abstract}
\affiliation{Department of Physics, Indian Institute of Technology, Hyderabad, Telangana-502284, India}
\maketitle
\section{Introduction}
The dark matter problem is one of the most vexing problems in Modern Physics and Astrophysics. Although dark matter constitutes about 25\% of total energy density of the universe and is a basic tenet of the $\Lambda$CDM model~\cite{Planck20}, its identity is still a mystery, despite close to 100 years of evidence~\cite{Hooper}. There is also no laboratory evidence for some of the most well motivated dark matter candidates such as WIMP or the axion or through indirect searches which have only reported null results~\cite{Merritt,Gaskins,Desai04,Boveia}.
Furthermore, the currently established $\Lambda$CDM model has also faced some tensions at scales smaller than 1 Mpc, such as the core/cusp problem, missing satellite problem, too big to fail problem, satellites plane problem etc~\cite{Bullock,Weinberg}. Data on galactic scales  for spiral galaxies have also revealed some intriguing deterministic scaling relations or correlations  such as Radial Acceleration Relation~\cite{McGaugh16},  constancy of dark matter halo surface density~\cite{Donato09}, which cannot be trivially reproduced using the  $\Lambda$CDM model. Therefore a large number of alternatives to the standard $\Lambda$CDM model have emerged such as Self-Interacting Dark Matter~\cite{SIDM}, Superfluid dark matter~\cite{Khoury}, Warm Dark Matter~\cite{Colin}, Wave (or Fuzzy) Dark Matter~\cite{EFerreira}, Flavor Mixed Dark Matter~\cite{Medvedev}, modified gravity which obviates the need for dark matter~\cite{LRR,Banik}, etc.

In order to ascertain if these small scales issues need a paradigm shift from $\Lambda$CDM or whether they can be explained using feedback processes from baryons, one should test the aforementioned relations with a plethora of dark matter dominated systems. We know that  Radial Acceleration relation does not hold for dwarf disk spirals, low surface brightness galaxies,  elliptical galaxies and galaxy groups~\cite{Salucci,ChanDesai,Gopika21}

Another class of objects to test if the above correlations are universal are galaxy clusters. Galaxy clusters are the most massive virialized objects
in the universe~\cite{Allen2011,Borgani12,Vikhlininrev}, which  have proved to be  wonderful laboratories for  galaxy evolution, Cosmology and Fundamental Physics~\cite{Allen2011,Vikhlininrev,Desai18,Boraalpha,BoraDesaiCDDR,BoraDM}. 
Galaxy clusters were the first astrophysical source which gave evidence for dark matter~\cite{Zwicky}. They have been shown to rule out  the universality of Radial Acceleration Relation~\cite{our_rar_2020,Tian,Chan20,Pradyumna21,Eckert22,ChanLaw} and constancy of halo surface density~\cite{Chan14,Gopika20,Gopika21}. However, some  scaling relations which are common between galaxies and galaxy clusters have also been found~\cite{Chandec22}.
Furthermore, galaxy clusters are known to exhibit many bivariate and  fundamental plane  scaling relations with tight scatter (See ~\cite{Maughan} for a recent review). Some of these include gas temperature vs total mass relation~\cite{Kravtsov06}, fundamental plane between characteristic radius, mass and temperature~\cite{Fujita,PradyumnaFP}, hot gas mass v/s total mass at intermediate radii~\cite{Chan2020}, etc.
Therefore, they continue to be very good laboratories for testing some of the anomalies and deterministic scaling relations obtained using spiral galaxies.

Most recently, Lovas (L22, hereafter)~\cite{Lovas} has found another intriguing property of dark matter haloes on galactic scales. Using 175 galaxies from the SPARC sample, L22 showed that the ratio of dark matter to baryonic matter $\left(\frac{M_{dark} (r)}{M_{bar} (r)}\right)$ shows a linear scaling as a function of radius. The residuals from the linear relation are small with $\sigma$=0.31, \rthis{where $\sigma$ was obtained  from the width of the best-fit Gaussian to the normalized  histogram of residuals from the linear relation.} When the radius is scaled using the radius ($r_1$) at which the aforementioned ratio is unity  ($r_{norm}= mr$) where $m=1/r_1$,  they find that $\left(\frac{M_{dark}}{M_{bar}}\right) = r_{norm}$.
L22 showed that this  linear trend was seen for  galaxies  of diverse morphologies in the SPARC sample from pressure supported to rotationally supported galaxies.  This linear scaling extends up to the last available kinematic data point~\cite{Lovas}. It is not immediately obvious whether the  linear trend for the ratio of dark matter to baryonic matter can be reproduced from the current theory of structure formation involving the $\Lambda$CDM model. In order to test whether this linear trend as a function of radius  is universal property of all dark matter haloes, we test this {\it ansatz} using  galaxy clusters from the HIFLUGCS sample~\cite{Chen}.

The outline of this manuscript is as follows.  The  HIFLUGCS cluster sample  as well as the ensuing analysis procedure  is described in Sect.~\ref{sec:dataset}. Our results are presented in Sect.~\ref{sec:results}. We conclude in Sect.~\ref{sec:conclusions}.
\section{Dataset  and Analysis}
\label{sec:dataset}
The HIFLUGCS cluster sample consists of 106 galaxy clusters and groups based on ROSAT and ASCA observations~\cite{Chen,Reiprich02}. Among these, 92 clusters have temperatures determined using X-ray spectroscopy whereas the remaining clusters had temperatures estimated using $L_x-T$ correlations obtained in Ref.~\cite{Markevitch}.  The X-ray surface brightness profiles have been  obtained for 36 clusters from the RASS survey and from pointed ROSAT observations for the remaining 70. The imaging has been done up to a radius 
of $r_{500}$. This final sample consists of about 52 cool core clusters and 54 non-cool core clusters. More details on the observations and data reductions can be found in Refs.~\cite{Chen,Reiprich02}.

The X-ray surface brightness profile was fit to both a single-$\beta$ and a double-$\beta$. The number density for a single-$\beta$ profile was was given by~\cite{betamodel}
\begin{equation}
n(r)=n_0\brak{1+\frac{r^2}{r_c^2}}^{-3\beta/2},
\label{eq:density}
\end{equation}
where $n_0$ is the central density, $r_c$ the core radius and $\beta$ is the index parameter. Assuming that the cluster gas is in hydrostatic equilibrium, the total cluster mass is given by~\cite{Chen}
\begin{equation}
M_{tot}(r) = \frac{3 \beta T_h r}{G \mu m_p} \frac{(r/r_c)^2}{1+(r/r_c)^2},
\label{eq:beta}
\end{equation}
where $\mu$ is the mean molecular weight equal to 0.59~\cite{Chen}. Eq.~\ref{eq:beta} assumes that the  gas is at constant temperature given by  $T_h$. This assumption is true for non-cool core clusters~\cite{Reiprich13}, where the temperature gradient in the hot gas is less than 8\%~\cite{Hudson10}.  However, the isothermality assumption  will not be applicable to the central regions in cool core clusters. Cool core clusters are also affected by AGN cooling and feedback near the center. Therefore, in our analysis, we only choose non-cool core clusters. All clusters with cooling time less than Hubble time (13 Gyr)  have been  identified as cool-core clusters and not used for this work. 

For the HIFLUGS sample, 49 clusters show lower $\chi^2$ with a double-$\beta$ model fit to the gas density profile. However,  no automated model selection techniques (such as AIC, BIC or Bayes factors) have been used to quantify the significance of goodness of fit of the double-$\beta$ profile with respect to single-$\beta$ model. \rthis{Therefore, it is hard to judge whether the double-$\beta$ profile is an improvement for these clusters.} Nevertheless, for this work we restrict our analysis to clusters for which single-$\beta$ profile provides the minimum $\chi^2$. The best-fit values of all the parameters \rthis{(along with the best-fit statistical errors obtained from the fit)} needed to estimate the hydrostatic mass from Eq.~\ref{eq:beta} are provided in Tables 1 and 2 of ~\cite{Chen}. Our final sample used for this work consists of 54 non-cool core clusters.

To test the linearity we need the total dark matter which can be obtained from the total  cluster mass as follows:
\begin{equation}
M_{DM}=M_{tot} (r) - M_{gas} (r)-M_{stars} (r),
\end{equation}
where $M_{gas}$ is the total gas mass and $M_{stars}$ is the total mass due to stars.
The gas mass  can be obtained by assuming spherical symmetry
\begin{equation}
    M_{gas} (r)=m_g \int 4\pi r^2 n(r') dr',
    \label{eq:eq3}
\end{equation}
where $n (r)$ is the gas density given in Eq.~\ref{eq:density}, $m_g$ is the average mass of each gas particle and is given by  $\mu_g m_p$, \rthis{where $\mu_g$=1.1548 is the mean molecular weight per electron~\cite{Bahar22}.} Although analytical expressions for the integral in Eq.~\ref{eq:eq3} have been  provided in literature~\cite{ChanLaw}, here we evaluate the integrals numerically to get the gas density at a given radius ($r$).  To obtain $M_{stars}$, we used the stellar to gas mass  scaling relations obtained in Ref.~\cite{Chiu18}.
This relation was derived using a sample of 91 SPT-SZ clusters detected up to a range of 1.3. This relation includes contribution from the BCG as well as cluster member galaxies within $r_{500}$. \rthis{We have also propagated the statistical uncertainty in the stellar mass to gas mass relation provided in Ref.~\cite{Chiu18}.}  The total baryonic mass ($M_{bar}$) is then given by the sum of gas and stellar mass
($M_{bar}=M_{gas}+M_{stars}$)

\section{Results}
\label{sec:results}
\subsection{Single cluster analysis}
\begin{figure}[H]
    \centering
    \subfigure[$\chi^2=144.8$]{\includegraphics[width=0.24\textwidth]{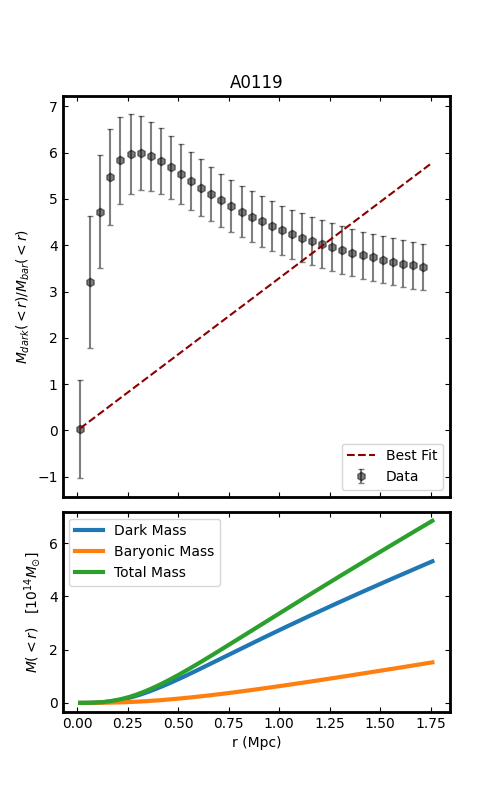}} 
    \subfigure[$\chi^2=90.0$]{\includegraphics[width=0.24\textwidth]{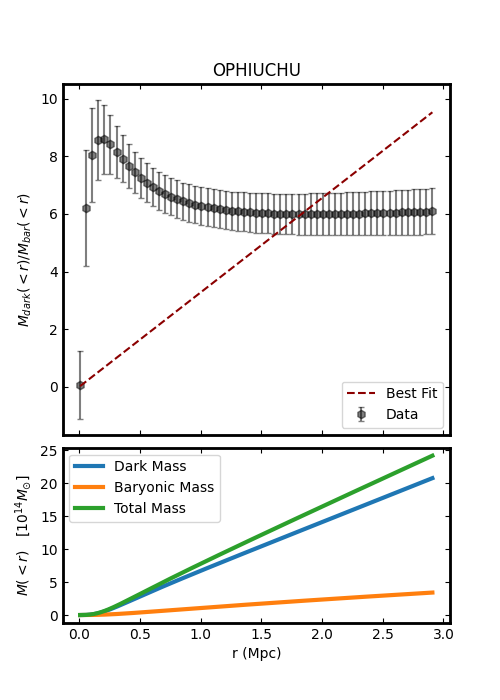}} 
    \subfigure[$\chi^2=59.9$]{\includegraphics[width=0.24\textwidth]{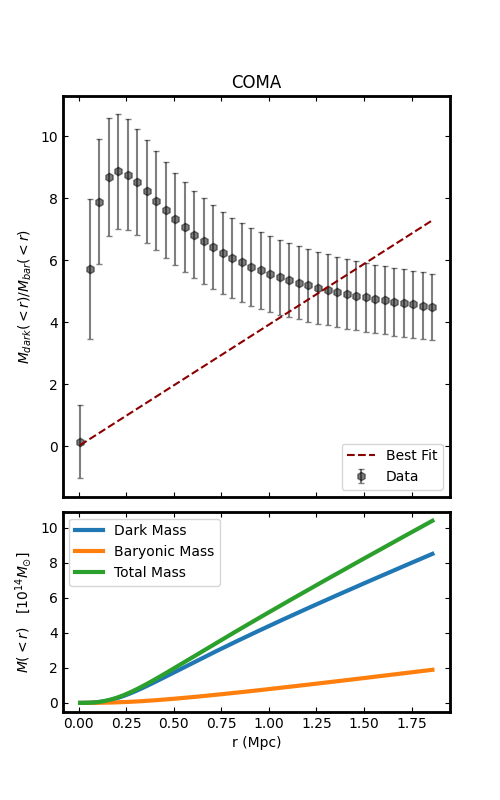}}
    \subfigure[$\chi^2=22.2$]{\includegraphics[width=0.24\textwidth]{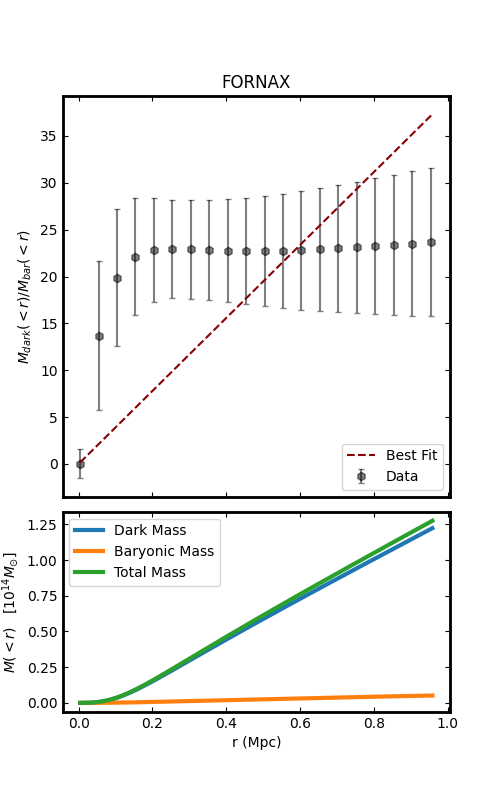}}
    \caption{The mass ratio and profiles for four clusters from the HIFLUGCS sample with large $\chi^2$ after fitting the ratio to linear function of the radius. One can compare these with similar plots for the SPARC sample in Fig. 2 of ~\cite{Lovas}, where one can see a linear trend.}
    \label{fig:clusters-fit}
\end{figure}

\begin{figure}[H]
    \centering
    \subfigure[$\chi^2=4.5$]{\includegraphics[width=0.24\textwidth]{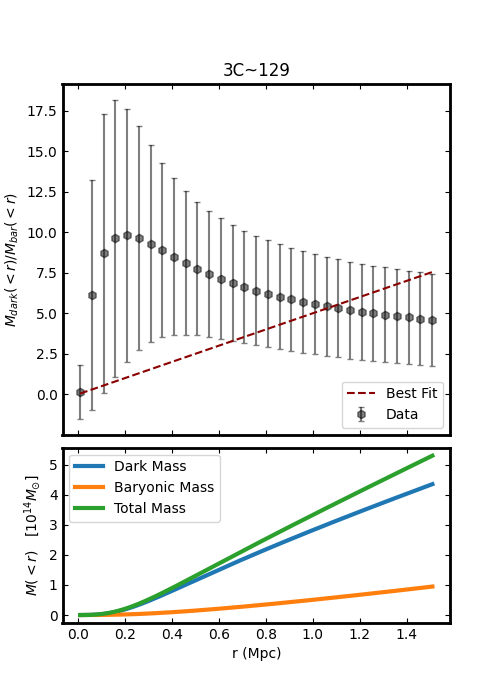}}
    \subfigure[$\chi^2=4.3$]{\includegraphics[width=0.24\textwidth]{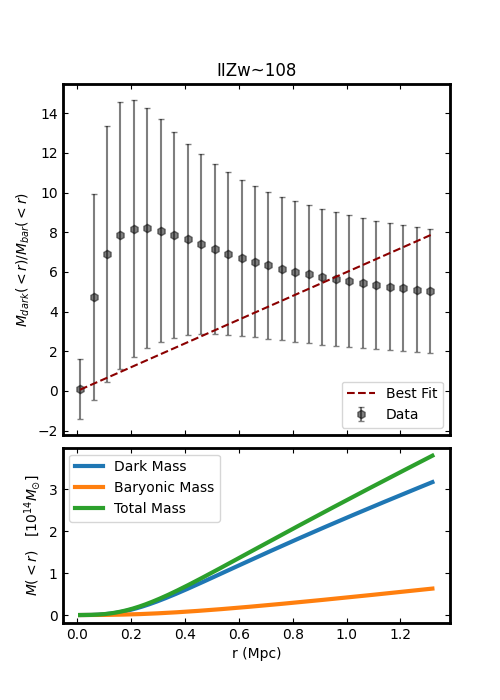}}
    \subfigure[$\chi^2=2.4$]{\includegraphics[width=0.24\textwidth]{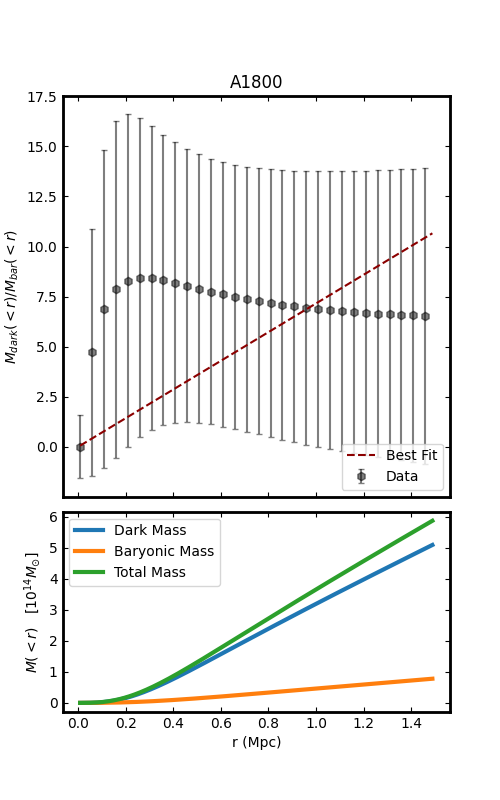}}    \subfigure[$\chi^2=2.7$]{\includegraphics[width=0.24\textwidth]{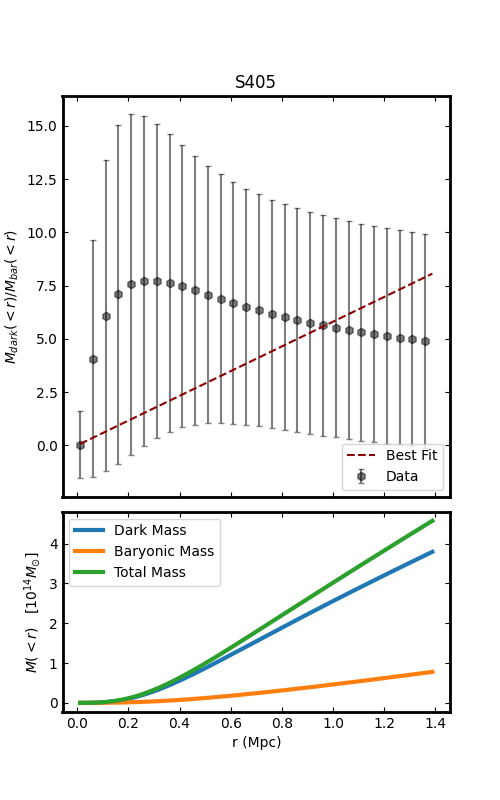}}
    \caption{The mass ratio and profiles for four clusters with high errors and $\chi^2/\text{dof}$ less than 1.}
    \label{fig:clusters-fit-err}
\end{figure}

We calculate the ratio of  dark matter to baryonic matter \rthis{mass} at 10 uniformly spaced values from $r=0$ to $r=r_{500}$. We then fit this ratio ($M_{DM}/M_{bar}$) to a  simple linear multiple of $r$ ($M_{DM}/M_{bar}=kr$) and obtain the best-fit value of $k$ using $\chi^2$ minimization as follows:
\begin{equation}
\chi^2= \sum  \left(\frac{M_{DM}/M_{bar}-kr}{\sigma_i}\right)^2,
\end{equation}
where $\sigma_i$ is the error in the ratio $M_{DM}/M_{bar}$ and is obtained by error propagation after taking into account the \rthis{statistical} uncertainties in $M_{bar}$, $M_{tot}$, and $M_{stars}$.  Since we are selecting 10 points per cluster with one free parameter, we have a total of 9 degrees of freedom (dof) \rthis{for all the clusters.} The best-fit $\chi^2$ values along with their $p$ values obtained from the $\chi^2$ cdf for each of the 54 non-cool core clusters can be found in Table~\ref{table:chi-p-clusters}. As we can see, most clusters (34) from our sample have $\chi^2$/dof \rthis{much} greater than one \rthis{and very small $p$-values.} \rthis{This shows that the linearity ansatz cannot be correct} and  the ratio of dark to baryonic matter \rthis{mass}  in galaxy clusters  \rthis{shows a non-linear trend} as a function of $r$, unlike the galaxies in the SPARC sample. A plot of this ratio as a function of $r$ can be found \rthis{for some sample clusters}  in Fig.~\ref{fig:clusters-fit}.  We however have about 20 clusters with $\chi^2/$dof$  \leq 1$. These are due to large observational errors in the ratio, \rthis{as opposed to the ratio showing a linear trend.} A plot  for some of these clusters with $\chi^2$/dof less than one can be found in Fig.~\ref{fig:clusters-fit-err}. However, even for these clusters with low $\chi^2$,  visually we do not see a discerning linear trend as a function of $r$.  

\def\arraystretch{1.26}
\begin{table}[H]
    % \centering
    \begin{tabular}{cccccccc}
    \toprule
        \textbf{Cluster Name} & \textbf{$\chi^2$}& \textbf{$\chi^2/$dof} & \textbf{p-value} & \textbf{Cluster Name} & \textbf{$\chi^2$} & \textbf{$\chi^2/$dof}& \textbf{p-value} \\ \toprule
A0119	&	144.8	&	16.09	&	0	&	MKW8	&	7.3	&	0.81	&	0.7	\\
A0399	&	10.3	&	1.14	&	0.42	&	ZwCl~1215	&	9.4	&	1.04	&	0.5	\\
A0400	&	163.8	&	18.2	&	0	&	3C~129	&	4.5	&	0.5	&	0.92	\\
A0401	&	248.5	&	27.61	&	0	&	A0548e	&	130.2	&	14.47	&	0	\\
A0576	&	2.4	&	0.27	&	0.99	&	A0548w	&	3.9	&	0.43	&	0.95	\\
A0754	&	143.4	&	15.93	&	0	&	A1775	&	61.8	&	6.87	&	0	\\
A1367	&	127.2	&	14.13	&	0	&	A1800	&	2.4	&	0.27	&	0.99	\\
A1644	&	7.6	&	0.84	&	0.67	&	A2319	&	195.3	&	21.7	&	0	\\
A1650	&	8.8	&	0.98	&	0.56	&	A2734	&	65.9	&	7.32	&	0	\\
A1736	&	5.6	&	0.62	&	0.85	&	A2877	&	9.6	&	1.07	&	0.48	\\
A2065	&	1.1	&	0.12	&	1	&	A3395n	&	1.6	&	0.18	&	1	\\
A2147	&	15.1	&	1.68	&	0.13	&	A3528n	&	73.4	&	8.16	&	0	\\
A2163	&	84.8	&	9.42	&	0	&	A3530	&	19.3	&	2.14	&	0.04	\\
A2255	&	70.5	&	7.83	&	0	&	A3532	&	50.9	&	5.66	&	0	\\
A2256	&	76.3	&	8.48	&	0	&	A3560	&	12.5	&	1.39	&	0.25	\\
A2634	&	49.2	&	5.47	&	0	&	A3627	&	39.4	&	4.38	&	0	\\
A3158	&	147.3	&	16.37	&	0	&	A3695	&	2.7	&	0.3	&	0.99	\\
A3266	&	240.5	&	26.72	&	0	&	A3822	&	6.4	&	0.71	&	0.78	\\
A3376	&	25.1	&	2.79	&	0.01	&	A3827	&	2	&	0.22	&	1	\\
A3391	&	61.9	&	6.88	&	0	&	A3888	&	6.2	&	0.69	&	0.79	\\
A3395s	&	5.3	&	0.59	&	0.87	&	A3921	&	88.1	&	9.79	&	0	\\
A3558	&	821	&	91.22	&	0	&	IIZw$\sim$108	&	4.3	&	0.48	&	0.93	\\
A3562	&	331.5	&	36.83	&	0	&	OPHIUCHU	&	90	&	10	&	0	\\
A3667	&	449.7	&	49.97	&	0	&	RXJ2344	&	8.6	&	0.96	&	0.57	\\
COMA	&	59.9	&	6.66	&	0	&	S405	&	2.7	&	0.3	&	0.99	\\
FORNAX	&	22.2	&	2.47	&	0.01	&	S636	&	5.8	&	0.64	&	0.83	\\
IIIZw54	&	2.7	&	0.3	&	0.99	&	TRIANGUL	&	548.3	&	60.92	&	0	\\
\toprule
    \end{tabular}
        \caption{\label{table:chi-p-clusters} Table showing $\chi^2$ for 9 degrees of freedom along with the $p$-value obtained using the $\chi^2$ cdf, while fitting the ratio of dark to baryonic matter to a linear relation as a function of radius.}
\end{table}

\subsection{Stacked Analysis}
\begin{figure}[H]
    \centering
    \includegraphics[width=0.7\textwidth]{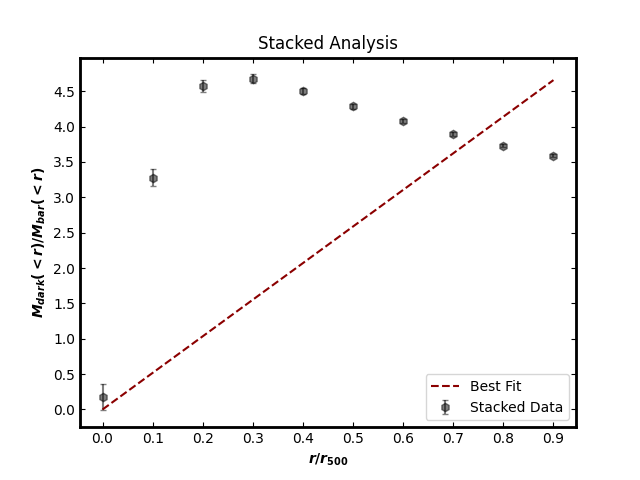}
    \caption{The stacked data points for all the clusters along with the "best" fit line. As we can see the fit is quite bad with best-fit $\chi^2$ equal to 8397.7 for 9 degrees of freedom.}
    \label{fig:stacked}
\end{figure}

In order to beat down the uncertainties  in the individual cluster analysis, we stack the data for all the non-cool core clusters.  \rthis{The stacking  of profiles from different clusters assumes that the profile is universal, which implies that  combining the data from  the  clusters in a consistent way would increase the significance of any putative correlation. Since clusters are expected to be  self-similar objects~\cite{Kaiser86}, this is a reasonable assumption even though the ratio in each cluster is non-linear.}  For the stacked analysis, we scale the radius by $r_{500}$ and calculate the ratio of dark to baryonic matter at 10 equally spaced values of $r/r_{500}$ between 0 and 1. The mean ratio $\bar{R}$  at each rescaled radius is given by the weighted mean 
\begin{equation}
\bar{R}= \frac{\sum \limits_{i=1}^N R_i/\sigma_i^2}{\sum \limits_{i=1}^N 1/\sigma_i^2},
\end{equation}
where $R_i$ and $\sigma_i$ are the associated values of the ratio and its uncertainty at each rescaled radius. The uncertainty in $\bar{R}$ 
($\sigma_R$) is given by
\begin{equation}
\sigma_R^2 = \frac{1}{\sum \limits_{i=1}^N 1/\sigma_i^2}
\end{equation}
We then stack the data for all the non-cool core clusters using 10 equally spaced values of $r/r_{500}$ from 0 to 1. This plot can be found in Fig.~\ref{fig:stacked}. As we can see, the stacked ratio does not obey a linear trend and we find appreciable deviations from a linear trend with  $\chi^2$/dof equal to 1512. Therefore, we find that the dark to baryonic matter ratio for the stacked sample does not show a linear trend.

\subsection{Discussion}
 \rthis{For our analysis, we have propagated the statistical errors in the best-fit parameters. We now discuss possible systematic effects which could affect our conclusions.  The main  assumption used to obtain the total cluster mass and thereby the dark matter mass is that of hydrostatic equilibrium. This involves a number of assumptions, one of them being that the gas pressure is purely thermal~\cite{Gianfagna21,Gianfagna22}. It has been shown using simulations that the hydrostatic mass can be underestimated with respect to  the true mass by 20-30\% due to non-hydrostatic pressure support~\cite{Gianfagna21}.  One would need to compare the X-ray masses with lensing masses to get a robust estimate of the hydrostatic bias in order to evaluate its impact on our underlying results.  Another assumption  we have made in  calculating the gas mass is to assume a constant temperature. Although this assumption is not true for cool-core clusters (which we have omitted), it should be sound for non cool-core clusters. However, it could introduce systematic effects up to 3\% in the estimation of central density~\cite{Chan14}. Finally, for some clusters there could be systematics due to  single-$\beta$ profile, instead of a double-$\beta$ profile.  The impact of this assumption could also be checked by redoing the analysis using the gas mass obtained by fitting the pressure profile to a double-$\beta$ profile.}

\rthis{Finally, another possible systematic could be due to the  effect of stellar mass. Over the years, many empirical relations have been used to model the stellar mass contribution of the total mass budget of clusters~\cite{Giodini09,Lin11,Andreon,Palmese}. However, the stellar mass  is only a small part of  the total baryonic mass budget, so using the stellar mass relations from Ref.~\cite{Chiu18} should not impact our results.}

\section{Conclusions}
 \label{sec:conclusions}
Recently, L22 found that the ratio of dark matter to baryonic matter shows a linear trend with radius upto the last measurable data point for a diverse class of galaxies in the SPARC sample with rms scatter of 0.31~\cite{Lovas}. It is not obvious whether this linear trend is a prediction of CDM haloes. In order to ascertain if this relation is a universal property of all dark matter haloes, we carry out a similar test for galaxy clusters. 

For this purpose we use a sample of 54 non-cool core clusters from the HIFLUGCS sample, which are in hydrostatic equilibrium. We used an isothermal  temperature profile and single-$\beta$ model for the gas density (cf. Eq.~\ref{eq:density}) to calculate the gas mass.  The total mass is obtained by assuming the clusters are in hydrostatic equilibrium (cf. Eq.~\ref{eq:beta}.) The stellar mass and gas mass were obtained using the scaling relations between the stellar and gas mass derived in ~\cite{Chiu18}, and by assuming a spherical symmetry (cf. Eq.~\ref{eq:eq3}) respectively. The dark mass is obtained by subtracting the gas and stellar masses from the total mass. Using this, we can calculate the ratio of dark to baryonic matter as function of radius from the cluster center up to $r_{500}$.

We then look for a linear trend in this ratio and quantifying any deviations using $\chi^2$. The best-fit $\chi^2$ value for each of the 54 clusters can be found in Table~\ref{table:chi-p-clusters}. As we can see most of the clusters have $\chi^2$/dof much greater than one, indicating that the ratio of the dark matter to baryonic matter does not scale linearly with radius. The plots for some of these clusters can be found in Fig.~\ref{fig:clusters-fit}. About 20 clusters have $\chi^2$/dof  less than  one. The ratio as a function of radius for some of these can be found in Fig.~\ref{fig:clusters-fit-err}. The low $\chi^2$ for these clusters is because of large error bars and by eye, we do not discern a linear trend. Therefore, none of the individual clusters show evidence for linearity in the ratio of dark to baryonic matter as a function of radius. Finally,  we stack the data as a function of radius scaled by $r_{500}$, and calculated the weighted  mean of this ratio for the entire cluster sample.   This ratio can be found in Fig.~\ref{fig:stacked}. We find that  the stacked ratio also does not scale linearly as a function of radius. \rthis{We have also discussed some possible systematic errors which we have not considered.}

Therefore, we conclude the the linear ratio for the dark matter to baryonic matter found by L22 for galaxies in the SPARC sample cannot be a universal property of all dark matter dominated systems, since galaxy clusters do not show a similar linear trend.

\section*{Acknowledgements}
\rthis{We are grateful to Man-Ho Chan for helping us correct a mistake in the first  draft of the manuscript. We also acknowledge the  anonymous referee for constructive feedback and comments on our manuscript.}
\bibliography{main}

\end{document}